Routines to synthesize carbyne of more than 6000 atoms

[1,2]Chi Ho Wong

[1]*Department of Industrial and Systems Engineering, The Hong Kong Polytechnic University, Hong Kong*

[2]*Research Institute for Advanced Manufacturing, The Hong Kong Polytechnic University, Hong Kong*

*Email: roych.wong@polyu.edu.hk*

Abstract:

The superior electronic, optical and magnetic properties of carbyne have been called for optoelectronic and magnetoelectronic applications. However, manufacturing a monoatomic chain of more than 6000 carbon atoms presents a huge technical challenge. In order to predict the optimal chain length in different environments, we develop a Monte Carlo model in which a finite-length carbyne in the size of 4000-15000 atoms is encapsulated by a carbon nanotube at finite temperatures. Our Monte Carlo simulation shows that the stability of the carbyne-nanotube is influenced by the charity and porosity of carbon nanotube, external pressure, temperature and the chain length…etc. When the geometric structure of carbon nanotube and environmental parameters are provided, our Monte Carlo algorithm can predict the maximum length of the internal carbyne. Our work presents a path to manufacture a carbon chain much longer than 6000 atoms at room temperature

1. Introduction

Effective manipulation of spin and charge in magnetic semiconductors is crucial to design robust spintronic devices [1-4]. Though novel properties of magnetic semiconductor are always emerged in nanoscale [5,6], the discovery of III-V semiconductor with the Curie transition above room temperature is rare [7]. To push the spintronic nano-devices at the next level, the edge magnetism [8] of semiconductors has to be investigated comprehensively. Carbyne is the one-dimensional monoatomic semiconductor consisting of sp-hybridized carbon atoms [9-15] where the ferromagnetism of pristine short carbon chains has been observed up to at least 400K [16]. Several dopants have been proposed to strengthen the magnetism of carbyne-based semiconductors with a local magnetic moment above 1.7$\mu_B$ [16,17]. The monoatomic structure makes carbyne the best candidate to study the physics of edge magnetism as it is much thinner than the nanowire in one-unit-cell thick (or an ultra-thin zigzag graphene stripe…etc[18,19]). The spin-spin interaction of a monoatomic chain contributes one of the edge magnetisms in a higher dimensional Bravais lattice. If the chain length of magnetic carbyne can be adjusted properly, the length-dependent spin-spin interaction down to monoatomic scale can be studied experimentally which not only gives valuable information to monitor the interfaces of magnetic junctions for spintronics [20], but also provides insight to the interface problems of magnetic heterostructures [21] at industrial level. Unlike the favourable kink structure in the short carbon chains [16,22,23], no magnetism has reported in the ~6000-atom-long carbon chain [24]. To emerge magnetism in a long carbon chain, the dopants from IV, V, and VII groups should be needed [16,17]. However, the dopants should be carefully introduced to avoid chain-breaking or self-dissociation. Unfortunately, there is no systematic routine to tune the chain length of the doped carbyne stably that impedes the bottom-up experimental studies of edge magnetism.

It is foreseeable that monitoring the stability of the doped carbyne will be much tougher than that of pristine carbyne in laboratories [17]. However, controlling the stability of pristine carbon chain is a stepping stone for manufacturing the doped carbyne in the future. The way to manufacture bulk carbyne is still an open question. Making a ~20-atom-long carbon chain had been a challenging problem over the past decade [13,25]. In order to extend the chain length, L.Shi et al encapsulated the carbyne by carbon nanotube (CNT) as a nanoreactor, which results in a carbon chain of up to ~6000 atoms [24]. Based on the analysis of the carbyne-nanotube, L Shi et al have argued that Van der Waal's VDW force from the carbon nanotube plays a major role to stabilize the internal carbyne [24,26]. Inspired by L.Shi et al's work, C.H.Wong et al developed a Monte Carlo method (LCC model) to study the stability of free-standing carbon chains laterally coupled by VDW force at finite temperatures [27]. In the absence of carbon nanotube, the direct exposure of carbyne to the environment may sacrifice the long chain-length characteristic but the kink structure in short carbyne may trigger magnetism [16]. One year later, the LCC model assisted C.H.Wong et al to manufacture a room-temperature ferromagnetic VDW-coupled free-standing carbon nanowire array in ~100-atom-long successfully [16]. This observation agrees with the L.Shi et al 's argument that the VDW interaction can be used to stabilize carbyne. Effect of substrate is critical to the stability of carbyne [28-30]. If copper substrate is used, the linear carbon chain can be extended to ~1000-atom-long [30].

Using carbon nanotube as a nanoreactor for sample fabrication towards bulk carbyne is a great discovery [24]. To lengthen the carbyne even longer, theoretical guidelines are needed. However, the DFT study of finite-length carbyne is only limited in ~20 atoms due to an unaffordable computational cost [31]. For example, running a ~6000-atom-long carbyne surrounded by CNT (over ~100000 atoms per unit cell) is always not a realistic task for DFT software [32]. To speed up the computational progress, using DFT software to compile an infinitely long LCC with a few atoms per unit cell is a favourite option unavoidably [17, 33].

DFT methods are ideal for situations at 0K [34]. DFT results are usually representative unless the materials are extremely unstable at finite temperatures. When a long linear carbon chain LCC dissociates itself at room temperature, a transfer function should be built on top of DFT methods in order to tune the theoretical results close to experimental observations. On the other hand, it is unclear how sample quality affects the growth of the internal LCC.

Concerning the above issues, we are going to develop a Monte Carlo algorithm to predict the stability of carbyne-nanotube at finite temperatures in a reasonable computational cost. A stochastic process is monitored to forecast the stability of the carbyne-nanotube up to 300K. Parametric studies of the composite are focused on external pressure, temperature, the chain length of carbyne, the chirality and the porosity of carbon nanotube…etc. In this work, we propose a path to stabilize an undoped carbyne that is a stepping stone to stabilize the doped carbyne for the bottom-up experimental studies of magnetism from monoatomic scale.

## 2. Computational methods:

### 2.1. Hamiltonian

The Hamiltonian of the carbyne-nanotube with length L is

$$H = e^{-T/T_{bj}}\left(\sum_{n=1,3,5}^{N}|E_{n,j}-E_1|e^{-\frac{\ell_n-\ell_{n,j}^{eq}}{0.5\ell_{n,j}^{eq}}} + \sum_{n=2,4,6}^{N}|E_{n,j'}-E_3|e^{-\frac{\ell_n-\ell_{n,j'}^{eq}}{0.5\ell_{n,j'}^{eq}}}\right) + e^{-T/T_{bj}}\sum_{n=1,2,3}^{N}J_A e^{-\frac{\ell_n-\ell_{n,j}^{eq}}{0.5\ell_{n,j}^{eq}}}(\cos\theta+1)^2$$

$$-4\varphi\sum_{\phi=0}^{2\pi}\sum_{n=1}^{N}\left[\left(\frac{\sigma}{r}\right)^6-\left(\frac{\sigma}{r}\right)^{12}\right]$$

where N is the total number of atoms in the carbon chain and T is the surrounding temperature. The $C-C$, $C=C$ and $C\equiv C$ bond energies are $E_1$ = 348kJ/mol, $E_2$ = 614 kJ/mol and $E_3$ = 839 kJ/mol, respectively [27]. The type of covalent bond is proposed by stochastic variables j and j'. For example, $E_{n,j}$(n = 100, j = 3) refers to the 100[th] carbon atom (n = 100) forming a triple bond (j = 3) with respect to the 99[th] carbon atom (n-1 = 99). The temperature of bond dissociation is $T_{bj} = E_j/k_B$ where $k_B$ is the Boltzmann constant [35]. The chain-stability factor is defined by $e^{-\frac{\ell_n-\ell_{n,j}^{eq}}{0.5\ell_{n,j}^{eq}}}$ [27].

The bond distance $\ell$ is computed in the Cartesian coordinate. The longitudinal axis refers to x-axis while y- and z-axis forms a lateral plane. The $C-C$, $C=C$ and $C\equiv C$ bond length on the ground state are $\ell_{n,1}^{eq}$ = 154pm, $\ell_{n,2}^{eq}$ = 134pm and $\ell_{n,3}^{eq}$ = 120pm, respectively [27]. For the kink term, the bond angle (or pivot angle) between three adjacent carbon atoms is θ. A linear carbon chain means θ = 180 degrees where the angular energy $J_A$ is 600 kJ/mol [27]. The Van der Waal's interaction between the carbon nanotube and the internal chain is

$E_{vdw} = -4\varphi\sum_{\phi=0}^{2\pi}\sum_{n=1}^{N}\left[\left(\frac{\sigma}{r}\right)^6-\left(\frac{\sigma}{r}\right)^{12}\right]$ [27]. The $\sum_{\phi=0}^{2\pi}$ sums up Van der Waal's interaction along the angular plane ϕ of CNT and r is the radial separation between LCC and CNT. By considering $L\frac{dE_{vdw}^2}{dL^2} = \frac{1}{\zeta}$ and $\frac{\partial E_{vdw}}{\partial r} = 0$, the calculated Van der Waal's constants are $\sigma \sim 1.2\times10^{-10}$m and $\varphi \sim 8\times10^{-23}$J [27] where $\zeta$ is the isothermal compressibility. Van der Waal's interaction is the only coupling between LCC and CNT.

### 2.2. Initial condition

A long LCC in 6000-atom-long (unless otherwise specified) surrounded by a single-walled carbon nanotube CNT is studied which is abbreviated as LCC@($N_c$,$M_c$)CNT. The LCC@($N_c$,$M_c$)CNT is spaced by ~0.5nm laterally where ($N_c$,$M_c$) refers to the chirality of CNT. After applying geometric optimization to a periodic LCC@($N_c$,$M_c$)CNT at the GGA-PBE level [36,37], the repeated unit of CNT is imported to our Monte Carlo simulation. The initial bond length of the internal LCC (cumulene phase: consecutive double bond) is 134pm. The lengths of LCC and ($N_c$,$M_c$)CNT are the same. The length of ($N_c$,$M_c$)CNT in our Monte Carlo simulation can be scaled with help of the repeated unit.

2.3. Algorithm

During the Monte Carlo iterations, the atomic coordinate and the type of covalent bond in the carbon chain are amended at finite temperatures. At each Monte Carlo step MCS, we select an atom in the carbon chain randomly and then calculate the initial Hamiltonian.

To propose the trial types of covalent bonds at each MCS, we assign a random number $0 \leq R_{bond} \leq 1$ in Table 1. The proposed types of covalent bond depend on the value of $R_{bond}$. For examples, the selected $n^{th}$ atom forming two double bonds with its nearest neighbours is expressed as $[=C=]$; If $[=C=]$ is detected and $R_{bond} = 0.8$, the trial type of covalent bond is $[-C-]$.

Table 1: The trial state of covalent bond.

| (A) If $[=C=]$ is detected | Trial state |
|---|---|
| $0 \leq R_{bond} < 0.33$ | $[\equiv C-]$ or $[-C\equiv]$ in equal probability |
| $0.33 \leq R_{bond} < 0.66$ | $[=C-]$ or $[-C=]$ in equal probability |
| $0.66 \leq R_{bond} \leq 1$ | $[-C-]$ |
| (B) If $[\equiv C-]$ or $[-C\equiv]$ is detected | Trial state |
| $0 \leq R_{bond} < 0.33$ | $[=C=]$ |
| $0.33 \leq R_{bond} < 0.66$ | $[=C-]$ or $[-C=]$ in equal probability |
| $0.66 \leq R_{bond} \leq 1$ | $[-C-]$ |
| (C) If $[-C-]$ is detected | Trial state |
| $0 \leq R_{bond} < 0.33$ | $[\equiv C-]$ or $[-C\equiv]$ in equal probability |
| $0.33 \leq R_{bond} < 0.66$ | $[=C-]$ or $[-C=]$ in equal probability |
| $0.66 \leq R_{bond} \leq 1$ | $[=C=]$ |
| (D) If $[=C-]$ or $[-C=]$ is detected | Trial state |
| $0 \leq R_{bond} < 0.33$ | $[\equiv C-]$ or $[-C\equiv]$ in equal probability |
| $0.33 \leq R_{bond} < 0.66$ | $[-C-]$ |
| $0.66 \leq R_{bond} \leq 1$ | $[=C=]$ |

Based on the metropolis algorithm [35], we propose a trial range of spatial fluctuation at each MCS. Three random numbers $R_x$, $R_y$, $R_z$ from 0 to 1 are generated, respectively. If $R_x > 0.5$, the selected atom moves to the +x direction longitudinally. Otherwise, it moves to -x direction. The same cut-off value, 0.5, is also applied to the sign convention along y- and z-axis, respectively. After the trial directions are proposed, the trial fluctuations ($\delta x$, $\delta y$, $\delta z$) of the carbon atom are $\delta x = \pm v <\delta t> R_c$ and $\delta y = \delta z \sim \dfrac{k_B T}{<E_1 + E_2 + E_3>} \delta x$, where the free particle velocity is $v = \sqrt{\dfrac{k_B T}{M}}$ [38]. The average scattering time $<\delta t>$ and the mass of a carbon atom M are 1.98x $10^{-13}$s and 19.9x$10^{-27}$kg, respectively [27]. The random number $0 < R_c < 1$ is to fine-tune the spatial fluctuation at each MCS.

The Hooke's factor, $f(Hooke) = \dfrac{K_{break}\left(\ell_{max} - <\ell_{eq}>\right)}{K_{carbyne@CNT}\left(\ell_{carbyne@CNT}\big|_{0K} - <\ell_{eq}>\right)}$, acts as a mean-free-path MFP controller to adjust the relative dynamics $\delta x \cdot f(Hooke), \delta y \cdot f(Hooke), \delta z \cdot f(Hooke)$ from a chain-breaking state to a crystalline form. The Hooke's factor mimics a situation that a stable carbyne nanotube should have a shorter MFP and has a narrower range of spatial fluctuation than an unstable one. The numerator of $f(Hooke)$ marks the elastic force when an isolated LCC is elongated from a ground state to a break point [38]. The break point refers to the formation of the maximum $C-C$ length $\ell_{max} \sim 1.73$ Å where the MFP is the largest [39]. The ground state of an isolated LCC (cumulene phase) has an average bond length of $<\ell_{eq}> \sim 1.34$Å. The spring constant of an LCC at the break point is $K_{break}$. The denominator of $f(Hooke)$ monitors the elastic force of a very long LCC protected by CNT relative to the same $<\ell_{eq}>$ [38]. The spring constant and the average bond length of the internal LCC at 0K is $K_{carbyne@CNT}$ and $\ell_{carbyne@CNT}\big|_{0K}$, respectively. In our DFT simulation, the GGA-PBE functional is selected [37]. By setting the bond length to be 0.173nm, the $K_{break}$ of an isolated infinitely long LCC chain is acquirable. The $K_{carbyne@CNT}$ refers to the spring constant of the internal LCC along the chain axis only.

After the trial range of spatial fluctuations and the trial types of covalent bond are proposed, we estimate the trial Hamiltonian. If the trial Hamiltonian is less positive (or more negative) than the initial Hamiltonian, the trial states are accepted [35]. Otherwise, the system returns to the previous states [35]. Thermal excitation is another opportunity to accept or reject the trial states in parallel. If the random number $0 \le R_B \le 1$ is smaller than the Boltzmann factor $e^{\frac{-\Delta H}{k_B T}}$, the trial states are accepted [35]. Otherwise, the system rejects the trial states. The Monte Carlo process will continue until equilibrium (i.e. ~250000 steps). We average the data in the range of $230000 \le MCS \le 250000$. We impose a boundary condition that the location of the 1st atom in LCC is fixed and meanwhile the covalent bond between the 1st and 2nd atoms in LCC is always $C-C$ bond. The initial condition reruns at each temperature. Octet rule [38] is applied.

2.3: Pre-calibration

The average bond distance of a finite length LCC in the CNT $\ell_{carbyne@CNT}\big|_{0K}$ is an unknown before the simulation starts. To guess a trial value of $\ell_{carbyne@CNT}\big|_{0K}$ reasonably, we use the recently announced LCC model [27] to generate a long LCC with length L. We calculate the average bond distance of LCC $\ell_{carbyne}\big|_{501K}$ above the Peierls transition temperature at ~500K [33] in which the polyyne phase forms automatically. As $K_{carbyne@CNT}$ and $<\ell_{eq}>$ refer to the situation at 0K, the $\ell_{carbyne}\big|_{0K}$ is estimated with the help of

$$\frac{\ell_{carbyne}\big|_{501K} - \ell_{carbyne}\big|_{0K}}{\ell_{carbyne}\big|_{0K}} = \alpha \Delta T$$, where the thermal expansion coefficient is $\alpha = 7 \times 10^{-5} K^{-1}$ [27,40]

To obtain the value of $\ell_{carbyne@CNT}\big|_{0K}$, we do a pre-calibration by substituting the trial $\ell_{carbyne}\big|_{0K}$ into the Hooke's factor. Then we run the Monte Carlo simulation (Section 2.2) of LCC@$(N_c, M_c)$CNT at T ~ 0K with the trial Hooke's factor. During the Monte Carlo iterations of the internal LCC, the effect of CNT tunes the average bond length from $\ell_{carbyne}\big|_{0K}$ to $\ell_{carbyne@CNT}\big|_{0K}$. After the pre-calibration, we substitute $\ell_{carbyne@CNT}\big|_{0K}$ to the Hooke's factor. With the calibrated Hooke's factor, our simulation tool is ready to study the properties of LCC@$(N_c, M_c)$CNT at finite temperatures.

3. Results:

Thermal excitation from 0K to 300K has changed the proportion of polyvne and cumulene phase in Figure 1. We have found that the probability of forming $C_{13}$ bond (polyvne phase) in the internal LCC decreases from 1.00 to 0.96 upon heating. Due to the growth of polyvne phase, the probability of forming $C_{22}$ bond (cumulene) increases to ~0.04 correspondingly. The formations of $C_{12}$ ($[=C-]$ or $[-C=]$) and $C_{11}$ ($[-C-]$) bond are rare. The most energetically favourable phase of the internal LCC is still polyvne even when temperature is at 300K [24]. After the pre-calibration process at T ~ 0K, the average bond length of the internal LCC is ~1.3Å.

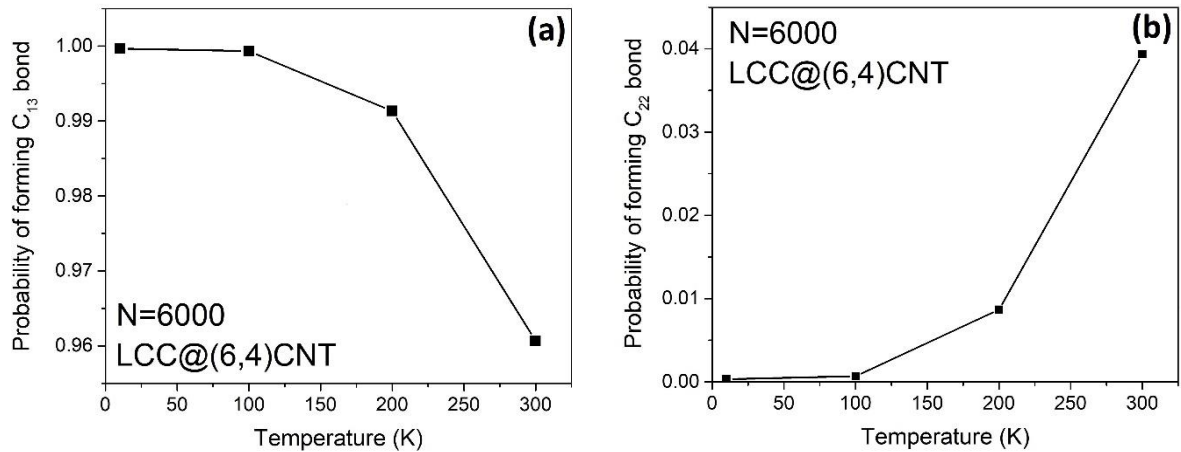

Figure 1: A LCC encapsulated by (6,4)CNT upon heating. The LCC is made up of 6000 atoms. **a** The probability of maintaining a polyvne phase in the carbon chain. The $C_{13}$ bond refers to the formation of single bond and triple bond alternatingly. **b** The probability of maintaining a cumulene phase in the carbon chain. The $C_{22}$ bond refers to the formation of consecutive double bonds.

As illustrated in Figure 2a, the stability of the carbyne-nanotube depends not only on the radius of CNT but also on chirality ($N_c$, $M_c$) [24,26]. In the carbyne-nanotube experiment [24], (6,4), (6,5) and (8,3)CNTs are the most abundant components. Under a constant temperature of 300K, our simulation data shows that (6,4)CNT should hold a longer LCC than (6,5)CNT. Although the experiment do not contain (5,0), (5,1), (6,1) and (5,4)CNTs, we run these samples in our simulation. The LCC@(5,1)CNT illustrate the highest chain-stability factor among the others. The chain-stability of LCC@(5,4)CNT is lower than that of LCC(6,4)CNT. The chain-stability of LCC@(5,0)CNT is the poorest. Figure 2b demonstrates how the kink angle of the internal LCC are affected by the geometry of CNT. The average kink angle of LCC@(5,0)CNT is much higher than the others. We observe that the higher the chain-stability factor is, the lower kink angle the LCC forms.

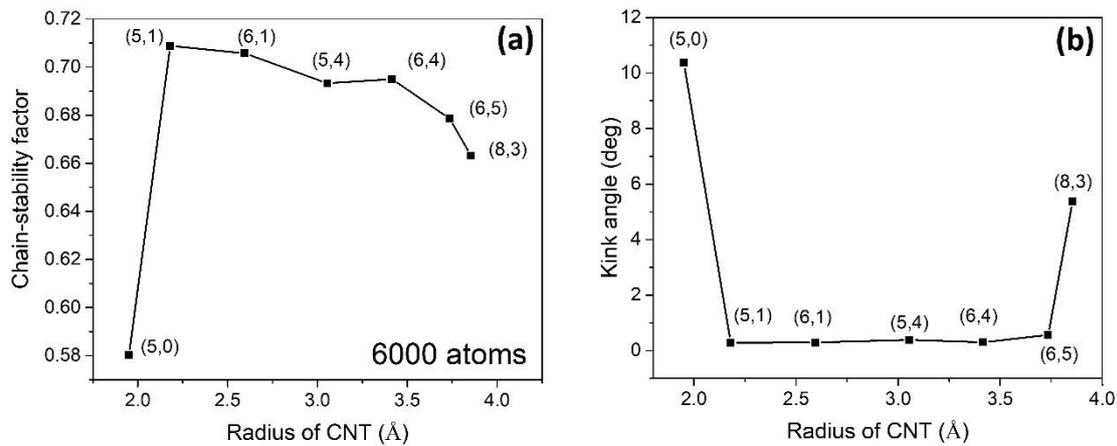

Figure 2: **a** The stability of the carbyne-nanotube depends on the radius of carbon nanotube $R_{CNT}$ and chirality ($N_c$, $M_c$). The surrounding temperature is 300K. **b** The average kink angles of the internal LCCs are plotted correspondingly. Zero kink angle (or pivot angle =180 deg) means that the chain is linear.

The chain stability of the internal LCC made up of 6000 atoms and 15000 atoms are both paled by thermal excitation in Figure 3. The 6000-atom-long LCC has a higher chain-stability factor than the 15000-atom-long LCC regardless of temperatures. By comparing the chain stabilities in both cases, we can interpret that the atomic fluctuation of the 6000-atom-long LCC is always lower at the same temperature. The chain-stability of the carbyne-nanotube in different lengths are compared in Figure 4. At T=300K, the chain-stability factor decreases gradually when N is smaller than 5500. However, for N > 6000, a rapid reduction in the chain-stability factor is observed where we bisect the cut-off length of LCC to be N = 5750. The inset of Figure 4 refers to the average energy of the internal LCC as a function of Monte Carlo iterations.

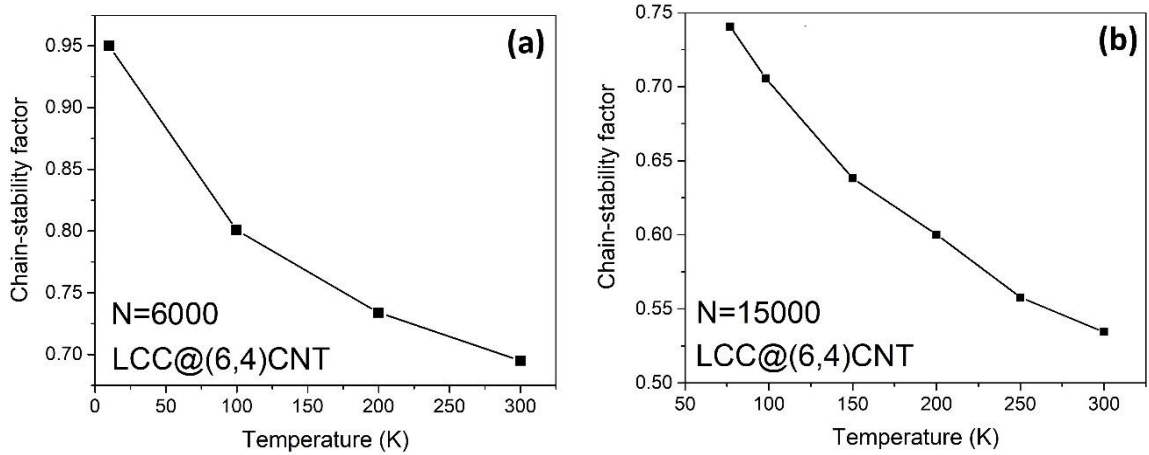

Figure 3: A LCC encapsulated by (6,4)CNT upon heating. The atomic fluctuation of LCC depends on the chain length and temperature. **a** The LCC consisted of 6000 atoms. **b** The LCC contains 15000 atoms.

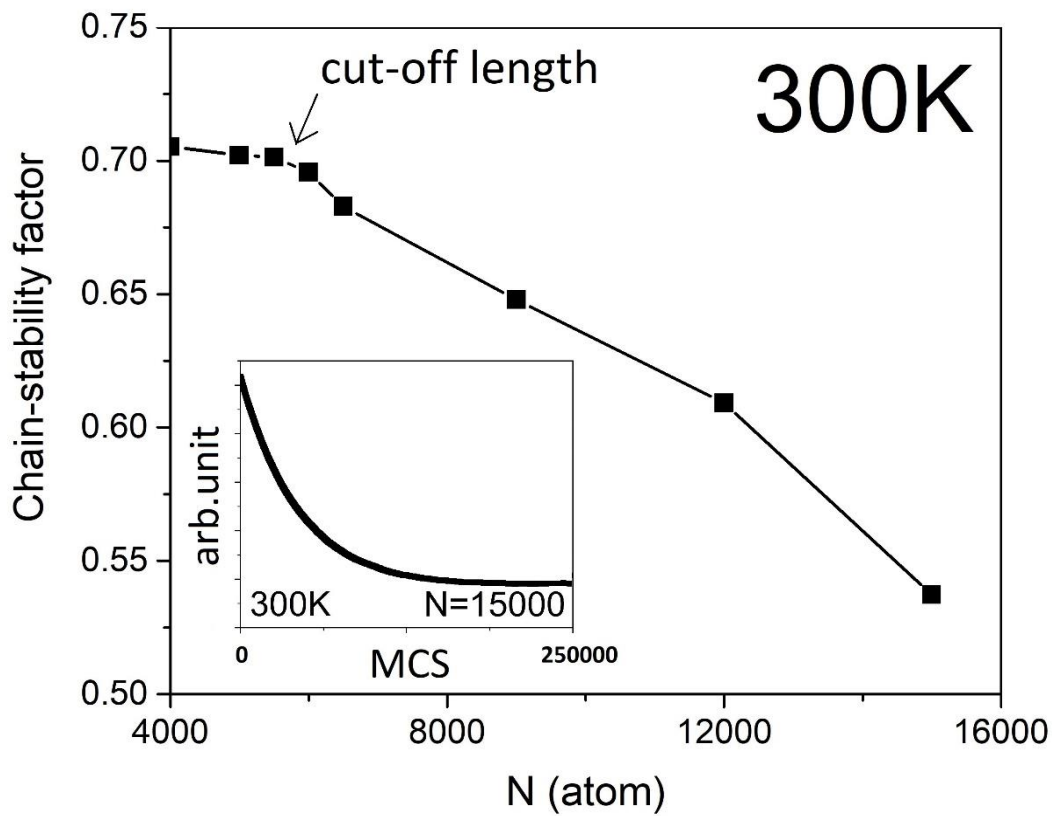

Figure 4: The chain-stability factor is a measure of atomic fluctuation of LCC@(6,4)CNT. The reduction in the chain-stability factor is more pronounced for N > 6000. The inset shows that our sample has reached an equilibrium state at the Monte Carlo steps above ~180000.

According to Table 2, applying compressive strain along the radial direction to LCC@(6,4)CNT pales the chain stability. The chain-stability factor drops from 0.69 to 0.650 under a compressive strain of 4%. The porosity of (6,4)CNT harms the chain-stability of the internal LCC with the evidence of Table 2. Introducing a tiny amount of vacancy defects is enough to shorten the cut-off length to N ~ 5600 significantly. We explore the cut-off length of LCC@($N_c$,$M_c$)CNT in various chiralities at 300K in Table 3. The internal LCC in (5,1)CNT is the longest. The ~7200-atom-long LCC in (5,1)CNT may lengthen the chain to ~10000nm. The internal LCC in (6,4)CNT is ~10% longer than that in (6,5)CNT. Based on the data in Table 4, it is possible to lengthen the internal LCC to 15000-atom-long. However, the sample should be cooled at T~120K. Our model expects that the internal LCC in (6,4)CNT should increase by less than 10% at ice temperature.

Table 2: Influence of radial compression on the stability of LCC@(6,4)CNT. A 6000-atom-long LCC is enclosed by (6,4) CNT at 300K.

| Compressive strain (%) | Chain-stability factor |
|---|---|
| 0 | 0.695 |
| 2 | 0.680 |
| 4 | 0.651 |

Table 3: Effect of vacancy defects on the stability of LCC@(6,4)CNT at 300K

| Vacancy Defect (%) | Cut-off length of LCC@(6,4)CNT |
|---|---|
| 0 | 5750-atom-long |
| 2.5 | 5600-atom-long |

Table 4: Prediction of the LCC lengths in different carbyne-nanotubes at 300K.

| Chirality (Nc,Mc) | Cut-off length of LCC@(Nc,Mc)CNT |
|---|---|
| (8,3) | 4850-atom-long |
| (6,5) | 5250-atom-long |
| (6,4) | 5750-atom-long |
| (5,1) | 7200-atom-long |

Table 5: The predicted cut-off lengths of LCC in (6,4)CNT at different temperatures

| Cut-off length of LCC | Operating temperature (K) |
|---|---|
| 5750-atom-long | 300 |
| 6600-atom-long | 273 |
| 15000-atom-long | 122 |

4.Discussions:

Although the cumulene phase should dominate in an isolated LCC [33], the polyvne phase is more energetically favourable in LCC@($N_c$,$M_c$)CNT [24]. According to Figure 1, the polyvne phase pales upon heating because the atomic fluctuation is more serious at high temperatures. This process can be controlled by the Boltzmann factor which accepts the trial type of covalent bond easier at high temperatures [27,35]. The formation of $[-C=]$, $[=C-]$ and $[-C-]$ are rare which confirms our Monte Carlo simulation monitoring the process of energy minimization effectively. For instances, the energy difference between $[-C\equiv]$ and $[=C=]$ is $|348 + 839 - (614 + 614)| = 41$kJ/mol per site. However, the energy difference (per site) between $[-C\equiv]$ and $[-C=]$, and the energy difference (per site) between $[-C\equiv]$ and $[-C-]$ are 225kJ/mol and 491kJ/mol per site, respectively. Hence, when the sample heats up, the LCC prefers to accept the trial state $[=C=]$ rather than $[-C=]$, $[=C-]$ and $[-C-]$.

We study the stability of the internal LCC as a function of $R_{CNT}$ in Figure 2a. While the chain length is fixed, the optimization process focuses on the radial direction of CNT only. An isolated long LCC is always unstable in a room-temperature environment owing to severe atomic fluctuations [10-15]. Kink angle is a measure of atomic fluctuation in LCC. The kink angle of LCC in Figure 2b can be used to judge whether the CNT provides an appropriate VDW protection or not. The average kink angle of the internal LCC in (5,1)CNT is just 0.3 degree (or pivot angle = 179.7 degrees) in Figure 2b. This explains why the LCC surrounded by (5,1)CNT shows the highest chain-stability factor where self-dissociation or chain-breaking is less likely to occur. Our simulation result is consistent to the experimental results that the internal LCC in (6,4)CNT is longer than that in (6,5)CNT. The chain-stability of LCC@(5,4)CNT is expected to be higher than that of LCC@(6,4)CNT. However, our simulation data shows an opposite behaviour. This is a proof that the stability of the internal LCC depends not only on the radius of CNT, but also on the chirality of CNT [13]. Mathematically, when compared to LCC(5,4)CNT, the (6,4) CNT suppresses the atomic fluctuations of LCC more effectively by ~4%.

The chain-stability factors of LCC@(6,4)CNT increase upon cooling in Figure 3 because the motion of carbon atoms are inactive at low temperatures [38]. If all covalent bonds in LCC are stable, the 15000-atom-long LCC should have weaker atomic fluctuations than the 6000-atom-long LCC. However, the comprehensive studies of LCC shows that lengthening LCC always cause chain-breaking or self-dissociation [10-15,24,26]. Up to our knowledge, forming all stable covalent bonds in an extremely long LCC have not been observed experimentally. To make a longer LCC unstable in our model, the Hooke's factor is used to tune the trial range of atomic fluctuations. When our previous LCC model is run this time, the maximum bond length of an isolated LCC calculated by the coefficient of thermal expansion is $\ell_{c-c} + \ell_{c-c}\alpha\Delta T \sim 1.58$Å. In this case, all the covalent bonds are stable and eventually $\ell_{carbyne}|_{0K}$ is shorter in a longer LCC. The accuracy of the coefficient of thermal expansion in our model has justified by Nathalia L et al's work [40]. After the $\ell_{carbyne}|_{0K}$ substituted into the pre-calibration process, the trial Hooke's factor increases in a longer LCC. By setting the maximum bond length in this carbyne-nanotube model to ~1.73Å [39], we impose a boundary condition that chain-breaking is not allowed even if two carbon atoms are spaced by ~1.73Å. Then our Monte Carlo system may force the atoms forming unstable covalent bonds at any chain length L. The unstable covalent bonds associated with large atomic fluctuations prevail the effect of stabilization towards bulk state. Therefore, the 15000-atom-long LCC (Figure 3b) has a lower chain-stability factor than the 6000-atom-long LCC (Figure 3a). If we do not pose this boundary condition, we cannot compare the stability of different types of LCC@($N_c$,$M_c$)CNT under the same length fairly.

Despite the Hooke's factor is designed semi-empirically, our simulation can mimic the stability of a long LCC that matches the experimental observation, where the cut-off length (5750-atom-long) in Figure 4 is comparable to the experimental data (~6000-atom-long) [24]. Indeed, the Hooke's factor is just a ratio of velocity because any elastic force multiplied by $<\delta t>/M$ gives a unit of velocity [38]. Therefore, $v \cdot f(Hooke)$ can be read as the adjustment of the trial velocity. If the bond length of the LCC in an extremely unstable carbyne-nanotube is ~1.73Å [39], the Hooke's factor approaches to 1 in which using the free-particle velocity as a trial velocity would be more appropriate [38]. The random number $R_c$ refines the spatial fluctuations to fit the process of energy minimization.

In our simulation, single-walled CNT is used to protect LCC instead of double-walled CNT. Using double-walled CNT should provide a better protection to the internal LCC [24,26]. Although our DFT simulation runs a parallel array of LCC@($N_c$,$M_c$)CNT laterally spaced by ~0.5nm, the lateral wall-to-wall interaction stabilizes the single-walled CNT in a good shape. The cut-off length of LCC@(6,4)CNT is slightly underestimated in our model. There are some sources of errors, the $\ell_{carbyne@CNT}|_{0K}$ in the Hooke's factor is obtained from the pre-calibration process for N > 4000 but the $K_{carbyne@CNT}$ is the spring constant of an infinitely long LCC. As the computational cost of DFT simulation in the size of over 4000 atoms per unit cell is not realistic [32], our model takes an approximation that the mechanical properties of LCC made up of more than 4000 atoms may move towards bulk state arguably [42]. Moreover, the iterative process of our Monte Carlo simulation does not apply to the atoms in CNT because the experimental results shows that the atomic vibration of CNT is much weaker that of the internal LCC. This experimental evidence allows us to take an approximation that the atoms in CNT are relatively at rest. Although using these approximations slightly underestimate the cut-off length, the computational cost becomes affordable. The inset of Figure 4 confirms that running 250000 Monte Carlo Steps is enough to drive the system into equilibrium even though the longest sample is chosen.

The shape of CNT plays a major role to stabilize the carbyne-nanotube. The compressive strain in Table 1 changes the CNT from a circular to elliptical shape. Although the elliptical CNT gives a slightly more negative VDW interactions along the minor axis (shorter axis), the VDW interactions along the major axis (longer axis) is much weaker. This is the reason why radial compression destabilizes LCC@(6,4)CNT. On the other hand, a stronger spatial fluctuation in the internal LCC is observed in porous CNT. When the selected atom in the internal LCC sums the negative VDW terms along the angular plane ϕ, the missing carbon atoms in the porous CNT fails to turns the VDW force more negatively and therefore maintaining a good sample quality of CNT is important to lengthen the internal LCC. We will discuss the VDW interaction more in the case of radial compression.

Our Monte Carlo results are consistent to several experiment observations such as (1) The LCC in (6,4)CNT more stable than the LCC in (6,5)CNT [24]; (2) The sample quality of

CNT affects the stability of the internal LCC [24,26]; (3) The cut-off length of LCC@(6,4)CNT is about ~6000-atom-long [24]. (4) Polyvne is the dominant phase in LCC@($N_c$,$M_c$)CNT at room-temperature [24]. These Monte Carlo results encourage us to predict the cut-off length of the internal LCC surrounded by different types of CNT in Table 3. Based on our Monte Carlo model, the LCC surrounded by (5,1)CNT should be the longest among the others because the atomic fluctuation is the weakest. The effect of free-radical electrons [38] in LCC is ignored in our model because the probability of forming polyvne phase is almost 1.

Here we unmask the process of energy minimization in our Monte Carlo system. Choosing an inappropriate type of covalent bond impedes the system to reach an equilibrium state. For example, when double bonds are proposed at the trial step, $|E_2 - E_1|$ and $|E_2 - E_3|$ will push the Hamiltonian more positive. Mathematically, the exponential term $e^{-\frac{\ell_n - \ell_{n,j}^{eq}}{0.5 \ell_{n,j}^{eq}}}$ is ~0.5 if $r_n$ is 1.73Å. This trial step sets $|E_2 - E_1| e^{-\frac{\ell_{\max} - \ell_{n,j}^{eq}}{0.5 \ell_{n,j}^{eq}}}$ and $|E_2 - E_3| e^{-\frac{\ell_{\max} - \ell_{n,j}^{eq}}{0.5 \ell_{n,j}^{eq}}}$ to ~130kJ/mol and ~110kJ/mol, respectively. Hence, double bond is usually rejected during the iterations, and the choices of single bond & triple bond are more preferable. If the polyvne phase is selected, $|E_1 - E_1| e^{-\frac{\ell_{\max} - \ell_{n,j}^{eq}}{0.5 \ell_{n,j}^{eq}}}$ and $|E_3 - E_3| e^{-\frac{\ell_{\max} - \ell_{n,j}^{eq}}{0.5 \ell_{n,j}^{eq}}}$ terms are vanished which makes the kink term and VDW term more significantly. As long as the polyvne phase is assigned, the adjustment of atomic coordinates relies only on the kink term and VDW terms. If the trial kink angle is large, the selected carbon atom experiences an uneven VDW interactions along the angular φ plane. The uneven VDW interaction make the Hamiltonian more positive which is consistent to the data in Table 1. For example, the selected carbon atom in LCC moved by 0.01nm orthogonally inside (5,4)CNT increases the VDW energy by 2% more positively. On the other hand, we apply DFT method to calculate the angular energy $E_{angular} \sim J_A (\cos\theta + 1)^2$ of a carbon chain at a pivot angle of 160 degrees in order to obtain the value of $J_A$. As the pivot angle smaller than 160 degrees (or kink angle > 20 degrees) is not noticeable in this project, our computed value of $J_A$ is applicable.

The radius of (5,0)CNT is 2nm which is still long enough to avoid the formation of new C-C bond between carbyne and CNT. For the chain-stability factors close to 0.7 in Figure 2, the average bond lengths of the LCCs are between ~1.3 and ~1.4A. However, the chain-stability factor of LCC@(5,0)CNT is as low as ~0.5 where the average bond lengths of the internal LCC is above 1.55Å. Unless the chain-length is massively shortened in (5,0)CNT, chain-breaking should occurs in a 6000-atom-long LCC@(5,0)CNT experimentally if the 6000 atoms are forced to connect together. Despite some carbyne-nanotubes are short, short carbyne nanotubes is still useful to tune the direct band gap of semiconductor continuously [41] and eventually controls the charge carriers across semiconductor junctions precisely, which may make a great impact to transistor technology.

The low computational cost of our Monte Carlo model is credited to two approximations. The first approximation is that the atomic positions of the finite-length CNT in our Monte Carlo model is obtained by scaling the repeating unit of CNT in the geometrically optimized infinitely long carbyne-nanotubes in the ab-initio calculation. This speeds up the computational progress because a time-consuming geometric optimization process of a big supercell (over 100000 atoms in a non-repeating unit) is avoidable. To validate this approximation, we consider the length dependence of the Raman spectrum of a finite-length CNT where the size effect starts to pale for a CNT length longer than ~200nm [43]. As the shape of the Raman spectrum is closely related to the atomic positions of materials and the minimum length of CNT in our model is far above 200nm, it is arguably that the first approximation will not create a large error in Monte Carlo results. The second approximation is that the Monte Carlo iterations are conducted to the internal chain only. This approximation can be validated by the experimental fact that the atomic fluctuations of CNT are relatively at rest when compared to the internal chain [24]. With the second approximation, the size of one-dimensional array in our Monte Carlo simulation under a C++ environment is 4000-15000 only, which reduces the computational time massively without creating a large error in Monte Carlo results. For example, any standalone high-performance PC can acquire one chain-stability datapoint of LCC@(6,4)CNT in a reasonable time: The ab-initio calculations with the first approximation spends about 2 days to finish, and the Monte Carlo simulation with the second approximation takes less than one hour to complete.

Conclusions:

Our Monte Carlo model is capable to the predict the cut-off length of LCC encapsulated by CNT at any temperature. The synthesis of the internal carbon chain of more than 6000 atoms is possible if the surrounding temperature and the type of CNT are chosen properly. The radius, the chirality and the sample quality of CNT give major impacts to the stabilization of the internal LCC. Radial compression on the carbyne-nanotube destabilize the internal LCC. Our model opens a new path to study the stability of carbyne in the size of 4000-15000 atoms at finite temperature which is a stepping stone to control the stability of the doped carbyne-nanotube for the bottom-up experimental studies of edge magnetism from monoatomic structure to a higher dimensional Bravais lattice.